\documentclass[a4paper,11pt]{article}
\pdfoutput=1
\usepackage{jheppub, lmodern, amsmath, amssymb, graphicx, xcolor, hyperref, slashed, multirow, rotating}

\usepackage{tikz}
\usepgflibrary{shapes.misc,shapes.geometric}
\usetikzlibrary{positioning}

\makeatletter\g@addto@macro\bfseries{\boldmath}\makeatother

\graphicspath{{figs/}}
\usepackage{latexsym}

\definecolor{commam}{rgb}{0.2,0.5,1.0}
\definecolor{myred}{rgb}{0.9, 0.3, 0.1}
\definecolor{myblue}{rgb}{0, 0, 0.7}
\definecolor{mygreen}{rgb}{0.04, 0.7, 0.5}

\hypersetup{colorlinks, citecolor=myred, linkcolor=myblue, urlcolor=myblue, linktocpage=true}
\def\figureautorefname~#1\null{Fig.\,#1\null}
\def\equationautorefname~#1\null{Eq.\,(#1)\null}

\def\be   {\begin{equation}}   \def\ee   {\end{equation}}
\def\ba   {\begin{array}}      \def\ea   {\end{array}}
\def\bea  {\begin{eqnarray}}   \def\eea  {\end{eqnarray}}
\def\bean {\begin{eqnarray*}}  \def\eean {\end{eqnarray*}}

\def\bry{\begin{array}}
\def\ery{\end{array}}

\newcommand{\FDF}[1][i]{\varphi^\dagger #1\!\overleftrightarrow{D}\!_\mu\varphi}
\newcommand{\FDFI}[1][i]{\varphi^\dagger #1\!\overleftrightarrow{D}^a\!\!\!_\mu\:\varphi}

\newcommand{\hc}[1]{#1}
\newcommand*{\tmp}[4]{\ensuremath{%
	{#4%
	\ifx\empty#3\empty\ifx\empty#1\empty\else^{#1}\fi\else^{#1(#3)}\fi%
	\ifx\empty#2\empty\else_{#2}\fi}%
}}

\newcommand*{\qq }[4][]{\tmp{#2}{#3}{#4}{#1{O}}}

\newcommand*{\ifb}{\ensuremath{\text{fb}^{-1}}}
\newcommand*{\iab}{\ensuremath{\text{ab}^{-1}}}
\newcommand*{\gev}{\ensuremath{\text{GeV}}}
\newcommand*{\tev}{\ensuremath{\text{TeV}}}

\newcommand*{\epem}{\ensuremath{e^+e^-}}
\newcommand*{\mpmm}{\ensuremath{\mu^+\mu^-}}
\newcommand*{\bbbar}{\ensuremath{b\,\bar b}}
\newcommand*{\ttbar}{\ensuremath{t\,\bar t}}
\newcommand*{\bwbw}{\ensuremath{bW^+\bar bW^-}}
\newcommand*{\eebb}{\ensuremath{\epem\to\bbbar}}
\newcommand*{\eett}{\ensuremath{\epem\to\ttbar}}

\DeclareMathAlphabet{\mathsfit}{\encodingdefault}{\sfdefault}{m}{sl}

\DeclareMathOperator{\cov}{cov}

\title{The top-quark window on compositeness\\at future lepton colliders}

\preprint{DESY 18-114}

\author{Gauthier Durieux,}
\author{Oleksii Matsedonskyi}
\affiliation{DESY Notkestra\ss e 85, D-22607, Hamburg, Germany}

\abstract{
In composite Higgs (CH) models, large mixings between the top quark and the new strongly interacting sector are required to generate its sizeable Yukawa coupling.
Precise measurements involving top as well as left-handed bottom quarks therefore offer an interesting opportunity to probe such new physics scenarios.
We study the impact of third-generation-quark pair production at future lepton colliders, translating prospective effective-field-theory sensitivities into the CH parameter space.
Our results show that one can probe a significant fraction of the natural CH parameter space through the top portal, especially at \tev\ centre-of-mass energies.  
}

\begin{document}

\maketitle

\section{Introduction}

Composite Higgs (CH) models~\cite{Kaplan:1983fs, Panico:2015jxa} represent attractive scenarios in which the gauge hierarchy problem is addressed by assuming that the Higgs boson is a composite bound state of a new strongly coupled dynamics. The Higgs boson potential then becomes insensitive to energies above the strong dynamics confinement scale. The sensitivity of the Higgs mass to the compositeness scale however requires the latter to lie not much higher than a few \tev. This motivates collider searches for various signatures of Higgs compositeness. Besides the solution they provide to the hierarchy problem, CH models can also address the dark matter puzzle~\cite{Fonseca:2015gva, Chala:2018qdf, Chala:2016ykx, Bruggisser:2016ixa, Balkin:2017yns}, flavour hierarchies~\cite{Barbieri:2012tu, Matsedonskyi:2014iha} and the matter-antimatter asymmetry observed in the universe~\cite{Espinosa:2011eu, Chala:2016ykx, Bruggisser:2018mrt, Bruggisser:2018mus}. 

One important characteristic of CH models is their two-sector structure~\cite{Contino:2006nn}. The field content of the elementary sector is analogous to that of the standard model (SM) but excludes the Higgs doublet. The latter, instead, belongs to the composite sector together with other composite states. The two sectors communicate through linear mass mixings between the elementary states and their composite partners. The mass eigenstates corresponding to SM particles thereby become partially composite. Higgs couplings with SM fermions and gauge bosons are also generated through this mechanism. In the case of fermions, mixings take the form
\be\label{eq:mix}
\epsilon_q m_\star \bar q Q + \epsilon_t m_\star \bar t T
\ee
where $q$, $t$ are respectively left- and right-handed SM fermions while $Q, T$ are their composite vector-like partners (with identical symmetry transformation properties for the fields of both chiralities). The mixing strengths are controlled by a typical strong sector mass scale $m_\star$ and suppressed by dimensionless parameters $\epsilon_q$ and $\epsilon_t$. They account for perturbations of the strong sector dynamics by the elementary fields and are typically expected to be at most of order one.

The composite vector-like partners have masses generated by the strong dynamics and of the order of $m_\star$. The mixings induce SM masses and open a portal between the SM fermions and the Higgs boson $\phi$. A simplistic way to introduce the Higgs boson couplings is through strong-sector Yukawa couplings:
\be\label{eq:yuk}
g_\star \bar Q \tilde \phi U
\ee
where $\tilde \phi_i = \epsilon_{ij} \phi^*_j$ and $g_\star$ is a typical coupling of the strong sector, expected to range between $1$ and $4\pi$. We then obtain see-saw-like expressions for the Yukawa couplings of SM fermions:
\be
\lambda_t  \simeq \frac{[q - Q \text{\,mixing}]}{[Q \text{\,mass}]} \times g_\star \times  \frac{[T-t\text{\,mixing}]}{[T \text{\,mass}]} 
\simeq  g_\star \epsilon_q \epsilon_t 
.
\ee 
The magnitudes of these Yukawa couplings are determined by the strength of the mixing of the corresponding elementary fields with their composite partners. The top quark, which has the largest Yukawa coupling, has the largest mixings and the strongest interactions with the composite sector. Since the left-handed bottom quark is tied to the left-handed top quark in a single $SU(2)_L$ doublet, it also inherits this large mixing. Studying top- and bottom-quark interactions with a high precision is therefore a powerful means to probe CH models. As new physics related to the hierarchy problem could lie in the several-\tev\ mass range, i.e.\ beyond the scales reachable through the direct production of heavy composite states at the LHC, indirect signals may be observed first.

In this paper, assuming null results in upcoming LHC resonance searches, we examine the potential of future lepton colliders to test the deviations predicted by natural CH models.
We focus primarily on top-quark pair production, and consider a new strong sector featuring a composite Higgs boson and linear mixing between the top-quark and new fermionic resonances. Other mixings, like that of the electron, are expected to be very small and hence negligible for our analysis. 
We describe the effects of the new strong sector below the compositeness scale
with an effective field theory (EFT) containing the SM states only, employing the so-called Warsaw basis of dimension-six operators~\cite{Grzadkowski:2010es}.
Examining the subset of operators affecting third-generation-quark pair production, we carefully estimate the magnitude of the various strong sector contributions to their coefficients.
The top-portal contributions, enhanced by the sizeable top-quark mixings to composite partners are explicitly shown to be dominant.
The sensitivity gained from top-quark pair production at future lepton colliders is then expressed in terms of the parameter space of CH models.  
In scenarios where the left-handed quark doublet of the third generation is significantly mixed with composite resonances, the high-energy production of bottom-quark pairs $\epem\to b\bar b$ also sets relevant constraints.
For comparison, we examine the complementary sensitivities brought by Higgs coupling measurements and universal contributions to $\epem\to\mu^+\mu^-$.
The relevance of the production of four third-generation quarks, $\epem\to t\bar{t}t\bar{t}$ or $t\bar{t}b\bar{b}$, remains to be examined.

\section{Quantifying Higgs compositeness effects}
\label{sec:quant}

To cover various concrete scenarios simultaneously, we adopt a model-independent approach to the description of the strong sector effects on top- and bottom-quark physics. 
An EFT involving SM states only is employed, with operators of dimension six at most.
The validity of this approach is ensured by our assumption that no resonance lies below the maximal centre-of-mass energy envisioned for the future lepton colliders that we consider. 
A discussion of \eett\ production in a specific CH model can be found in Ref.\,\cite{Barducci:2015aoa}.
In this section, we present the rules used to estimate the magnitudes of the effective operators coefficients generated by the strong sector, and discuss how these rules are affected by a particular choice for the EFT operator basis.

\subsection{Power counting rules}

We rely on three ingredients for estimating the magnitudes of the different operator coefficients: partial compositeness, dimensional analysis, and selection rules. 
Let us derive the associated power counting rules.

Partial compositeness requires SM fermionic field appearing in operators generated by the strong sector to be accompanied by the corresponding mixing factor $\epsilon_{q,t}$. The product of the left- and right-handed fermion mixings with the strong coupling $g_\star$ is fixed by the Yukawa coupling of the corresponding fermion, e.g.\ $\lambda_t \simeq g_\star \epsilon_q \epsilon_t$.

Dimensional analysis allows to determine the powers of couplings and mass parameters appearing in operator coefficients, by matching the operator dimensions to that of the action density. In natural units $\hbar=1=c$, the field and derivative content of an operator only fixes the energy dimension of its coefficient. Powers of couplings can also be determined once dimensions of mass $M$, length $L$ and time $T$ are restored (see e.g.~\cite{Panico:2015jxa}). 
The action density then has dimensions of $[\hbar][L]^{-4}$ where $[\hbar]=[M][L]^2/[T]$. 
It has to be matched by the overall dimensions of the various factors entering an effective operator. 
The length and $\hbar$ dimensions are respectively $-1$ and $1/2$ for scalar and vector fields, $-3/2$ and $1/2$ for fermions, $-1$ and $0$ for mass parameters and derivatives, $0$ and $-1/2$ for gauge and Yukawa couplings, $0$ and $1$ for $\hbar/(4\pi)^2$ loop factors.

Following Ref.~\cite{Giudice:2007fh}, we assume that all the strong sector effects can be characterized by a mass parameter $m_\star$ and a coupling $g_\star$.
Dimensional analysis would then lead to a $m_\star^4/g_\star^2$ estimate for the coefficient of an operator generated by the composite sector with no field insertion.
Every additional operator component then has to be accompanied by the appropriate powers of $g_\star$ and $m_\star$ compensating for its length and $\hbar$ dimensions.
For instance, each fermionic field $\psi$ comes with a factor of $g_\star/m_\star^{3/2}$, the Higgs doublet with a factor of $g_\star/m_\star$, and each derivative with a factor of $1/m_\star$. 
Since SM gauge fields $X_\mu$ only appear through covariant derivatives, $D_\mu = \partial_\mu + i g_X X_\mu$ where $g_X$ is the corresponding gauge coupling, they should be accompanied by a factor of $g_X/m_\star$.

The general form of an effective operator satisfying the principles of 
partial compositeness and dimensional analysis is therefore
\be\label{eq:powcount}
\frac{m_\star^4}{g_\star^2} \:\hat{O} \left(\epsilon_\psi \frac{g_\star \psi}{m_\star^{3/2}},  \frac{g_\star \phi }{m_\star}, \frac{\partial_\mu}{m_\star}, \frac{g_X X_\mu}{m_\star} \right) 
\ee
where $\hat{O}$ is a dimensionless function of its arguments. Dimensional-analysis estimates for operator coefficients can be corrected by dimensionless factors generically expected to be of order one. Selection rules can however lead to parametric suppressions, forcing for instance the appearance of additional loop and mixing factors.

\subsection{Operator basis reduction}
\label{sec:opbr}

Having established the power counting rules used to estimate the magnitude of operator coefficients, we now need to construct the operators themselves. Generically, one can expect that an effective theory obtained after integrating out the heavy composite sector could contain all operators allowed by symmetries, with coefficients following the rules established above.
Not all these operators are however independent. It is thus practical to reduce this redundant set to a basis including no redundant operators.
Standard techniques like integration by parts and field redefinitions can be employed. 
Field redefinitions can serve to effectively impose the SM equations of motion (\emph{eom}s) order-by-order in the EFT expansion. A dimension-six operator $\qq{}{i}{}$ of coefficient $c_i$ can then be traded for others:
\be
c_1 \qq{}{1}{} \to  \delta c_2\; \qq{}{2}{} + \cdots .
\ee 
\emph{Eom}s are for instance commonly used to re-express operators with more derivatives as combinations of operators with more fields.
If a Lagrangian containing redundant operators initially satisfies a given power counting, there however is no guarantee that this will still be the case after reduction of that set to a basis of independent operators.
Operator substitutions only preserve the power counting when the corrections they induce to operator coefficients do not exceed the initial power-counting estimates
\be\label{eq:pcbr}
\delta c_i \lesssim c_i
\,.
\ee

Since integration by parts does not change the field content of an operator, it will not break our power counting rules. On the other hand, the applications of certain \emph{eom}s can lead to violation of the condition \eqref{eq:pcbr}. Let us examine them one by one. For left-handed fermions, the \textit{eom}
\bea
\slashed D q = - i \lambda_t \tilde \phi t + \cdots
\eea
can be used to make the replacement
\be
\cdots \left[\frac{\slashed D}{m_\star}\right] \left[ \epsilon_q \frac{g_\star q}{m_\star^{3/2}} \right] \to 
\cdots\; 
	\left\{
	\epsilon_q^2\;\left[\frac{ g_\star \tilde \phi}{m_\star}\right] \left[\epsilon_t \frac{g_\star t}{m_\star^{3/2}}\right]
	+\cdots\right\}
\ee
where we used $\lambda_t\simeq g_\star \epsilon_q \epsilon_t$.
In comparison with the prescription~\eqref{eq:powcount}, the obtained operator is further suppressed by a factor of $\epsilon_q^2 \lesssim 1$. Therefore, the condition~\eqref{eq:pcbr} holds and the power counting is preserved by this replacement. The same conclusion is reached with the \textit{eom}s of the right-handed fermions.

The Higgs field {\it eom} reads
\bea
(D_\mu D^\mu \phi)^i = \mu^2 \phi^i - \lambda |\phi|^2 \phi^i + \lambda_t \epsilon^{ij} \bar q_{j} t + \cdots.
\eea
Following the same reasoning as for fermions, and using the smallness of the Higgs mass parameter $\mu^2 \ll m_\star^2$ and the quartic $\lambda \ll g_\star^2$, together with $\lambda_t \simeq g_\star \epsilon_q \epsilon_t$, we arrive at the conclusion that operators featuring a $D_\mu D^\mu \phi$ factor can be traded for others by applying the {\it eom} for $\phi$ without violating our power counting for operator coefficients:
\be
\cdots	\left[\frac{D}{m_\star}\right]^2 
	\left[\frac{g_\star \phi}{m_\star} \right]
\to\cdots\left\{
	\left.\frac{\mu^2}{m_\star^2}\right.
	\left[\frac{g_\star \phi}{m_\star}\right]
-
	\left.\frac{\lambda}{g_\star^2}\right.
	\left[\frac{g_\star \phi}{m_\star}\right]^3
+
	\left[\epsilon_q \frac{g_\star q}{m_\star^{3/2}}\right]
	\left[\epsilon_t \frac{g_\star t}{m_\star^{3/2}}\right]
+ \cdots
\right\}\,.
\ee

Finally, applying the \emph{eom}s of the weak gauge bosons
\be\label{eq:Weom}
(D^\mu W_{\mu \nu})^a = \frac{g}{2} ( \phi^\dagger i \overleftrightarrow D_\nu^a \phi +  \bar l \gamma_\nu \tau^a l + \bar q \gamma_\nu \tau^a q)
\ee
where
$
\phi^\dagger i\overleftrightarrow{D}_{\mu}^a \phi = \phi^\dagger (\tau^a iD_\mu \phi) + (\tau^a iD_\mu \phi)^\dagger \phi
$
and $\tau^a$ are the Pauli matrices, leads to:
\be
\cdots \left[\frac{D^\mu}{m_\star}\right] \left[\frac{g W_{\mu \nu}}{m_\star^2}\right] \to
\cdots  \left\{
	  \frac{g^2}{2g_\star^2}
		\left[\frac{g_\star\phi}{m_\star}\right]^2
		\left[\frac{D}{m_\star}\right]
	+ \frac {g^2} {2 g_\star^2\epsilon_l^2} \left[\epsilon_l\frac{g_\star l}{m_\star^{3/2}}\right]^2
	+ \frac {g^2} {2 g_\star^2 \epsilon_q^2} \left[\epsilon_q \frac{g_\star q}{m_\star^{3/2}}\right]^2
	\right\}.
\ee
While the operators generated by the first term above with additional Higgs fields do satisfy the condition \eqref{eq:pcbr} since $g\ll g_\star$, violations of the power counting rules can occur for operators generated with additional fermions, if the factors $g^2/g_\star^2 \epsilon^2_{q,l}$ exceed unity. This violation can be especially large for electrons which are constrained to have small mixings with composite resonances. 
Indeed, given that $\lambda_e\simeq g_\star \epsilon_l \epsilon_e$, we can estimate the typical electron mixing to be $\epsilon_l \simeq \epsilon_e \simeq \sqrt{\lambda_e/g_\star} \ll 1$.\footnote{The Yukawa coupling of the electron only fixes the product of left and right mixings. Neither of them is nevertheless allowed to be sizeable experimentally, as significant modifications of the electron couplings would be generated otherwise.}
The same observations can be made when applying the \emph{eom}s of other SM gauge bosons.

In conclusion, one can therefore not take any non-redundant basis of dimension-six operators and simply apply the power counting described in the previous section to obtain correct estimates of the low-energy effects of the new strong sector. A more careful treatment is required, tracing back the various redundant-operator contributions to each independent operator of interest, in order not to miss important effects. In particular, one needs to make sure to identify the contributions to any operator of phenomenological interest which is generated by application of the \emph{eom}s of the SM gauge fields.

\section{Future lepton collider sensitivities}

Having at our disposal power counting rules, we now proceed further to our goal: estimating the reach on CH models through processes sensitive to the top-quark mixings to the strong sector.
In this section, we identify all the operators of the the Warsaw basis affecting \ttbar\ production at lepton colliders with coefficients receiving contributions enhanced by top-quark mixings.
On the way, we also identify the universal contributions ---~independent of top-quark mixings~--- they receive and show that they are generically subdominant. Note that we do not aim at analysing here all the universal contributions to third-generation quark production. Other processes are arguably more sensitive to those.
We examine separately the sensitivity to various classes of operators, leaving a combined analysis and a discussion of its implications for CH models for the next section. 
As we demonstrate in \autoref{sec:disc}, the CH parameter space accessible via the processes which have the best sensitivity to the universal effects is complementary to the one accessible via the top portal.

\subsection{Operators relevant for \texorpdfstring{\eett, \bbbar}{e+ e- -> t t, b b}}
\label{sec:relevant_operators}

We start by identifying the operators of the Warsaw basis~\cite{Grzadkowski:2010es} affecting third-generation-quark production and generated by the strong sector with coefficients enhanced by the top-quark mixings.
As argued in \autoref{sec:opbr}, the contributions arising from operators eliminated from the Warsaw basis through the use of SM gauge field \emph{eom}s also have to be taken into account.

Warsaw-basis operators with coefficients enhanced by the top-quark mixings always contain top and the left-handed bottom quarks. Applying the \emph{eom}s to eliminate covariant derivatives does indeed not remove third-generation currents.
Operators containing top and left-handed bottom quarks which affect \eett, \bbbar\  production modify top- and left-handed bottom-quark interactions with the photon and $Z$ boson, or involve two third-generation quarks and two electrons.
We do not include operators of the Warsaw basis which do not receive contributions proportional to the top-quark mixing. They are expected to be better constrained through other measurements. The impact of all dominant universal contributions to operators containing top and left-handed bottom quarks will however be discussed.
We also discard chirality-breaking four fermion operators like $\bar{l}e\: \bar{q}u$, all contributions to their coefficients are suppressed by the electron mixing.
We thus focus on the following subset of operators from the Warsaw basis:
\begin{equation}
\let\varphi\phi
\begin{aligned}[m]
	\qq{1}{\varphi q}{ij}
	&=(\FDF) (\bar{q}_i\gamma^\mu q_j)
	,\\
	\qq{3}{\varphi q}{ij}
	&=(\FDFI) (\bar{q}_i\gamma^\mu\tau^a q_j)
	,\\
	\qq{}{\varphi u}{ij}
	&=(\FDF) (\bar{u}_i\gamma^\mu u_j)
	,\\
	\hc{\qq{}{uW}{ij}}
	&=(\bar{q}_i\sigma^{\mu\nu}\tau^au_j)\:\tilde{\varphi}W_{\mu\nu}^a
	,\\
	\hc{\qq{}{uB}{ij}}
	&=(\bar{q}_i\sigma^{\mu\nu} u_j)\quad\:\tilde{\varphi}B_{\mu\nu}
	,\\
\end{aligned}
\hspace{2cm}
\begin{aligned}[m]
	\qq{1}{lq}{ijkl}
	&=(\bar l_i\gamma^\mu l_j)
	  (\bar q_k\gamma^\mu q_l)
	,\\
	\qq{3}{lq}{ijkl}
	&=(\bar l_i\gamma^\mu \tau^a l_j)
	  (\bar q_k\gamma^\mu \tau^a q_l)
	,\\
	\qq{}{lu}{ijkl}
	&=(\bar l_i\gamma^\mu l_j)
	  (\bar u_k\gamma^\mu u_l)
	,\\
	\qq{}{eq}{ijkl}
	&=(\bar e_i\gamma^\mu e_j)
	  (\bar q_k\gamma^\mu q_l)
	,\\
	\qq{}{eu}{ijkl}
	&=(\bar e_i\gamma^\mu e_j)
	  (\bar u_k\gamma^\mu u_l)
	,\\
\end{aligned}
\label{eq:warsaw}
\end{equation}
where $q,l$ are left-handed quark and lepton doublets; $u,d,e$ are right-handed up-type, down-type quarks, and lepton singlets; $\phi^\dagger i\overleftrightarrow{D}_{\mu} \phi = \phi^\dagger (iD_\mu \phi) + (iD_\mu \phi)^\dagger \phi$; and $ijkl$ are generation indices.
For our purpose, generation indices should naturally be set to $3$ for quarks and $1$ for leptons. These flavour assignments will be implicitly assumed in the following.
It should be noted that only the $\qq{-}{\phi q}{}\equiv (\qq{1}{\phi q}{}-\qq{3}{\phi q}{})/2$ and $\qq{-}{lq}{}\equiv (\qq{1}{lq}{} - \qq{3}{lq}{})/2$ combinations of operators actually contain top quarks (and electrons) while the $\qq{+}{\phi q}{}$ and $\qq{+}{lq}{}$ combinations defined analogously contain bottom quarks (or neutrinos). Measurements of $\eett$ and $\eebb$ are therefore sensitive to independent combinations of operators featuring only $SU(2)_L$ doublets.

The coefficients of the operators above can be generated directly from the strong sector. As argued in the previous section, they can also obtain larger indirect contributions from redundant operators having been eliminated using the \emph{eom}s of the SM gauge fields. We list below all such redundant operators. They either involve a third-generation-quark current or are universal:
\begin{equation}\label{eq:redundop}
\begin{aligned}
	\qq{3}{qD}{} &= (\bar{q}\gamma^\mu \tau^a q) (D^\nu W^a_{\mu\nu}),\\
	\qq{1}{qD}{} &= (\bar{q}\gamma^\mu q) (D^\nu B_{\mu\nu}),\\
	\qq{}{uD}{}  &= (\bar{u}\gamma^\mu u) (D^\nu B_{\mu\nu}),
\end{aligned}\hspace{1cm}
\begin{aligned}
	\qq{}{2W}{} &= (D_\rho W^{a\mu\rho}) (D^\nu W^a_{\mu\nu}),\\
	\qq{}{2B}{} &= (D_\rho B^{\mu\rho}) (D^\nu B_{\mu\nu}),\\
	\qq{}{W}{} &= (\FDFI) (D_\nu W^{a\mu\nu}),\\
	\qq{}{B}{} &= (\FDF) (D_\nu B^{\mu\nu}).\\
\end{aligned}
\end{equation}
Applying the \emph{eom}s
\begin{align*}
(D_\nu W^{\nu\mu})^{a} &= \frac{g}{2}\left(
	\FDFI
	+ \bar q\gamma^\mu \tau^a q
	+ \bar l\gamma^\mu \tau^a l
	\right) ,\\\displaybreak[1]
D_\nu B^{\nu\mu} &= g'\left(
	\frac{1}{2} \FDF
	+\frac{1}{6} \bar q\gamma^\mu q
	+\frac{2}{3} \bar u\gamma^\mu u
	-\frac{1}{3} \bar d\gamma^\mu d
	-\frac{1}{2} \bar l\gamma^\mu l
	-\bar e\gamma^\mu e 
	\right)\,,
\end{align*}
the following contributions to the Warsaw basis operator coefficients are generated:
\begin{equation}
\begin{aligned}
\qq{3}{qD}{} &\to \frac{g}{2}(
	  \qq{3}{\phi q}{}
	+ \qq{3}{lq}{}
	+ \cdots
),\\
\qq{1}{qD}{} &\to \frac{g'}{2}(
	  \qq{1}{\phi q}{}
	- \qq{1}{l q}{}
	-2\qq{}{eq}{}
	+ \cdots
),\\
\qq{}{uD}{} &\to \frac{g'}{2}(
	  \qq{}{\phi u}{}
	- \qq{}{lu}{}
	+2\qq{}{eu}{}
	+ \cdots
),\\[2mm]
\qq{}{2W}{} &\to \frac{g^2}{2}(
	  \qq{3}{\phi q}{}
	+ \qq{3}{lq}{}
	+ \cdots
),\\
\qq{}{2B}{} &\to \frac{g'^2}{6}(
	  \qq{1}{\phi q}{}
	- \qq{1}{lq}{}
	-2 \qq{}{eq}{}
	+4\qq{}{\phi u}{}
	-4\qq{}{lu}{}
	-8\qq{}{eu}{}
	+ \cdots
),\\
\qq{}{W}{} &\to \frac{g}{2}(
	  \qq{3}{\phi q}{}
	+ \cdots
),\\
\qq{}{B}{} &\to \frac{g'}{6}(
	  \qq{1}{\phi q}{}
	+4\qq{}{\phi u}{}
	+ \cdots
).
\\[-1.5ex]
\end{aligned}
\label{eq:op_eoms}
\end{equation}

The removed universal operators also affect Warsaw-basis operators having no impact on top- and bottom-quark physics. Without aiming at a comprehensive study, we for instance note that operators involving four leptons will be generated when eliminating the redundant $\qq{}{2W,2B}{}$ operators:
\begin{equation}
\begin{aligned}
\qq{}{2W}{}	&\to \cdots + \frac{g^2}{2}
	(\bar{l}\gamma^\mu \tau^I l)(\bar{l}\gamma_\mu \tau^I l),
	\\
\qq{}{2B}{}	&\to \cdots + g'^2
	\left[
	\frac{1}{2}\: (\bar{l}\gamma^\mu l)(\bar{l}\gamma_\mu l)
	+(\bar{l}\gamma^\mu l)(\bar{e}\gamma_\mu e)	
	+2\:(\bar{e}\gamma^\mu e)(\bar{e}\gamma_\mu e)
	\right].
\end{aligned}
\end{equation}
To illustrate this connection we will later show the sensitivities induced by the measurement of the $\epem\to\mpmm$ process at multi-\tev\ centre-of-mass energies, next to those of \eett, \bbbar. We will however not cover precision electroweak measurements which would become relevant when such high-energy runs are not available.

\begin{sidewaystable}\centering
\newcommand{\coi}[1]{\tikz[remember picture]\coordinate(#1i) at (0,0);}
\newcommand{\cof}[1]{\tikz[remember picture]\coordinate(#1f) at (0,0);}
\newcommand{\co}[1]{\tikz[remember picture]\coordinate(#1) at (0,0);}
\newcommand{\myar}[1]{\draw[ar] (#1i)--(#1)-- (#1f);}
\renewcommand{\frac}[2]{#1/#2}
\renewcommand{\arraystretch}{1.4}
\ensuremath{\displaystyle\begin{array}{@{}l@{\hspace{.5cm}} l*{7}{@{}l} @{\hspace{.5cm}}l *{2}{c@{}}c@{\qquad}*{4}{c@{}}
@{\hspace{.75cm}}c@{\hspace{.5cm}}c
@{}c@{}
}
			&\multicolumn{8}{@{}c}{\text{power counting}}
			& \multicolumn{1}{@{}c}{\text{extra suppr.}} 
			& \multicolumn{7}{c}{\text{competitive indirect contributions}}
& \text{fully composite $u$}
& \text{equally composite $q$,$u$}
\\
\qq{1}{\phi q}{}&&&&\epsilon_q^2&&&g_\star^2&/m_\star^2&\multirow{2}{*}{\Bigg\}\text{anticorr.}}&\cdot&g'/2&\cdot&\cdot&g'^2/6&\cdot&g'/6&\lambda_t^2/m_\star^2&\lambda_t g_\star /m_\star^2&\\
\qq{3}{\phi q}{}&&&&\epsilon_q^2&&&g_\star^2&/m_\star^2&&g/2&\cdot&\cdot&g^2/2&\cdot&g/2&\cdot&\lambda_t^2/m_\star^2&\lambda_t g_\star /m_\star^2&\\
\qq{}{\phi u}{}&&&\epsilon_t^2&&&&g_\star^2&/m_\star^2&\epsilon_q^2&\cdot&\cdot&g'/2&\cdot&2g'^2/3&\cdot&2g'/3&\lambda_t^2/m_\star^2&\lambda_t^2 /m_\star^2&\\[3mm]
\qq{}{uW}{}&&&\epsilon_t&\epsilon_q&g&&g_\star&/m_\star^2&\frac{g_\star^2}{(4\pi)^2}&\cdot&\cdot&\cdot&\cdot&\cdot&\cdot&\cdot&\lambda_t g (g_\star/4\pi)^2/m_\star^2&\lambda_t g (g_\star/4\pi)^2/m_\star^2&\\
\qq{}{uB}{}&&&\epsilon_t&\epsilon_q&&g'&g_\star&/m_\star^2&\frac{g_\star^2}{(4\pi)^2}&\cdot&\cdot&\cdot&\cdot&\cdot&\cdot&\cdot&\lambda_t g' (g_\star/4\pi)^2/m_\star^2&\lambda_t g' (g_\star/4\pi)^2 /m_\star^2&\\[3mm]
\qq{1}{lq}{}&&\epsilon_l^2&&\epsilon_q^2&&&g_\star^2&/m_\star^2&&\cdot&-g'/2&\cdot&\cdot&-g'^2/6&\cdot&\cdot&\lambda_t^2g'^2/g_\star^2 m_\star^2&\lambda_t g'^2/g_\star m_\star^2&\\
\qq{3}{lq}{}&&\epsilon_l^2&&\epsilon_q^2&&&g_\star^2&/m_\star^2&&g/2&\cdot&\cdot&g^2/2&\cdot&\cdot&\cdot&\lambda_t^2g^2/g_\star^2 m_\star^2&\lambda_t g^2/g_\star m_\star^2&\\
\qq{}{lu}{}&&\epsilon_l^2&\epsilon_t^2&&&&g_\star^2&/m_\star^2&&\cdot&\cdot&-g'/2&\cdot&-2g'^2/3&\cdot&\cdot&g'^2/m_\star^2&\lambda_t g'^2/g_\star m_\star^2&\\
\qq{}{eq}{}&\epsilon_e^2&&&\epsilon_q^2&&&g_\star^2&/m_\star^2&&\cdot&-g'&\cdot&\cdot&-g'^2/3&\cdot&\cdot&\lambda_t^2g'^2/g_\star^2 m_\star^2&\lambda_t g'^2/g_\star m_\star^2&\\
\qq{}{eu}{}&\epsilon_e^2&&\epsilon_t^2&&&&g_\star^2&/m_\star^2&&\cdot&\cdot&-g'&\cdot&-4g'^2/3&\cdot&\cdot&g'^2/m_\star^2&\lambda_t g'^2/g_\star m_\star^2&

\\
&&&&&&&&&
&\cof{qD3}
&\cof{qD1}
&\cof{uD}
&\cof{DD3}
&\cof{DD1}
&\cof{pD3}
&\cof{pD1}
\\
\qq{3}{qD}{}		&&&&\epsilon_q^2&g&&&/m_\star^2
	&\coi{qD3}
	&\co{qD3}\\
\qq{1}{qD}{}		&&&&\epsilon_q^2&&g'&&/m_\star^2
	&\coi{qD1}
	&&\co{qD1}\\
\qq{}{uD}{}		&&&\epsilon_t^2&&&g'&&/m_\star^2
	&\coi{uD}
	&&&\co{uD}\\[5mm]
\qq{}{2W}{}		&&&&&g^2&&/g_\star^2&/m_\star^2
	&\coi{DD3}
	&&&&\co{DD3}\\
\qq{}{2B}{}		&&&&&&g'^2&/g_\star^2&/m_\star^2
	&\coi{DD1}
	&&&&&\co{DD1}\\[3mm]
\qq{}{W}{}	&&&&&g&&&/m_\star^2
	&\coi{pD3}
	&&&&&&\co{pD3}\\
\qq{}{B}{}	&&&&&&g'&&/m_\star^2
	&\coi{pD1}
	&&&&&&&\co{pD1}
\end{array}}
\begin{tikzpicture}[
	remember picture, overlay,
	ar/.style = {->, shorten >=0pt, shorten <=0pt,>=stealth, semithick}]
	\myar{qD3}
	\myar{qD1}
	\myar{uD}
	\myar{DD3}
	\myar{DD1}
	\myar{pD3}
	\myar{pD1}
\end{tikzpicture}
\caption{First column: Direct application of the power-counting rules to estimate the coefficients of Warsaw-basis operators contributing to third-generation quark production and enhanced by top-quark mixings. Redundant operators from which they receive contributions, after application of the \emph{eom}s for the electroweak gauge fields, are listed too. Second column: Extra suppression arising from selection rules. Third column: Additional factors affecting the indirect contributions of redundant operators to Warsaw-basis operators after application of the \emph{eom}s. Fourth and fifth columns: Dominant contributions in the case of a fully composite $t$ ($\epsilon_t=1$, $\epsilon_q=\lambda_t/g_\star$) and of equally composite $q$ and $t$ ($\epsilon_q=\epsilon_t = \sqrt{\lambda_t/g_\star}$). Subdominant contributions are suppressed by powers of $g^{(\prime)}/g_\star$, $g^{(\prime)}/\lambda_t$, or $\epsilon_{l,e}\;g_\star/g^{(\prime)}$. Numerical prefactors have been omitted in these last two columns.
}
\label{tab:coeffs}
\end{sidewaystable}

\subsection{Power-counting estimates}
\label{sec:power_counting_estimates}
A direct application of the power counting of \autoref{eq:powcount} for the operators in both \autoref{eq:warsaw} and \autoref{eq:redundop} leads to the estimates of \autoref{tab:coeffs} for their coefficients. In few cases, additional suppressions are necessary:
\begin{itemize}
\item For $\qq{1,3}{\phi q}{}$ operators, one often assumes a specific structure of the bottom quark mixing with the composite sector with a so-called $P_{LR}$ symmetry~\cite{Agashe:2006at}. As a result, the leading contributions of these two operators are perfectly anticorrelated and do not contribute to the coupling of the $Z$ boson to left-handed bottom quarks which is tightly constrained experimentally. A correlation between the left-handed $tbW$ and $ttZ$ couplings arising from the same $\qq{1,3}{\phi q}{}$ operators is then also induced~\cite{delAguila:2000aa, Aguilar-Saavedra:2013pxa, Grojean:2013qca}.

\item The correction to the coupling of the right-handed top quark to the $Z$ boson originating from the $\qq{}{\phi u}{}$ operator is also typically forbidden by the $P_{LR}$ symmetry. This time, the symmetry protection rather arises as an accident in minimal CH models~\cite{Grojean:2013qca}. It is broken by the left-handed top-quark mixing $\epsilon_q$, so that an estimate for the coefficient of this operator involves an additional $\epsilon_q^2$ suppression.

Such additional suppressions however only affect the zero-momentum corrections to the $Z$ boson couplings, and do not apply to $\qq{1,3}{uD,qD}{}$ operators.

\item The dipole $\qq{}{uB}{}$, $\qq{}{uW}{}$ operators always suffer from an additional $(g_\star/4 \pi)^2$ loop suppression in known UV complete CH models (see e.g. discussion in Refs.~\cite{Liu:2016idz,Chala:2017sjk}).
\end{itemize}

Indirect contributions arising through the replacements of \autoref{eq:op_eoms} which derive from the application of the electroweak gauge field \emph{eom}s are also displayed in \autoref{tab:coeffs}.
They always dominate over the direct ones for four-fermion operators involving leptons, given that $\epsilon_{l,e} \ll g^{(\prime)}/g_\star$. 
Moreover, among the indirect contributions, those of the $\qq{1,3}{qD,uD}{}$ operators are larger than those of the universal $\qq{}{2W,2B}{}$ since $g^{(\prime)} /g_\star \epsilon_{q,t} \simeq g^{(\prime)} \epsilon_{t,q}/\lambda_t\lesssim 1$.
On the other hand, for $\qq{1,3}{\phi q, \phi u}{}$ operators, the direct contributions are dominant compared to the indirect ones arising from $\qq{1,3}{qD,uD}{}$, $\qq{}{2W,2B}{}$ and $\qq{}{W,B}{}$ operators since $g^{(\prime)}<g_\star$,  $\;g^{(\prime)2}/\epsilon_{q,t}g_\star^2\simeq g^{(\prime)2}\epsilon_{t,q}/\lambda_t g_\star \lesssim 1$ and $g^{(\prime)} /g_\star \epsilon_{q,t} \simeq g^{(\prime)}\epsilon_{t,q}/\lambda_t \lesssim 1$, respectively.
An important conclusion from the discussion above is that the universal contributions ---~not depending on the top-quark mixings~--- to the coefficients of operators involving top and left-handed bottom quarks are always subdominant. 
The top-portal hence constitutes an exclusive probe compositeness.

We consider in the following
 two representative benchmark scenarios in which the $\epsilon_{q,t}$ mixings are fixed, leaving only $g_\star$ and $m_\star$ as free CH parameters. In the first case, the right-handed top quark is assumed to be fully composite so that $\epsilon_t=1$ and $\epsilon_q = \lambda_t/g_\star$~\cite{DeSimone:2012fs}. In the second scenario, the left- and right-handed top quarks are assumed to be equally composite so that $\epsilon_q = \epsilon_t = \sqrt{\lambda_t/g_\star}$. The dominant power-counting contributions to each Warsaw-basis operator in these scenarios are displayed in the last two columns of \autoref{tab:coeffs}.

\subsection{Sensitivities}
We now discuss the sensitivity of future lepton colliders.
CLIC-, ILC- and circular collider (CC)-like benchmark run scenarios are adopted. They are characterized by the centre-of-mass energies, luminosities and beam polarizations shown in \autoref{tab:run_scenarios}.
The CLIC-like scenario is directly taken from Table\,7 of Ref.\,\cite{CLIC:2016zwp}, omitting only the collection of $100\,\ifb$ forecast at $\sqrt{s}=350\,\gev$.
Our ILC-like scenario is freely inspired from the various ones discussed in Ref.\,\cite{Barklow:2015tja}. A $1\,\tev$ run is preferred over a luminosity upgrade at lower centre-of-mass energies since our focus is on top-quark pair production. 
According to Ref.\,\cite{Benedikt:2018}, the FCC-ee could gather $1.5\,\iab$ of integrated luminosity at a centre-of-mass energy of $365\,\gev$, in addition to $200\,\ifb$ at the top-quark pair production threshold, over a period of six years.
We adopt these numbers in our CC-like scenario.

\begin{table}\centering
\begin{tabular*}{\textwidth}{@{\extracolsep{\fill}} r*{7}{c}cr}
$\sqrt{s}$ [\tev]
& 0.35	& 0.365	& 0.38	& 0.5	& 1	& 1.4	& 3	& $P(e^+,e^-)$\\[1mm]\hline\noalign{\vskip1mm}
\multirow{3}{*}{$\mathcal{L}$ [\iab]}
&	&	& 0.5	&	&	& 1.5	& 3	& $(0,\mp 0.8)$		&CLIC 	\\
&	&	&	& 0.5	& 1	&	&	& $(\pm 0.3,\mp 0.8)$	&ILC	\\
& 0.2	& 1.5	&	&	& 	&	&	& $(0,0)$		&CC		
\\[1mm]\hline\noalign{\vskip1mm}
$\epsilon_{t\bar{t}}$ [\%]
& 10	& 10	& 10	& 10	& 6	& 6	& 5	\\
$\epsilon_{b\bar{b}}$ [\%]
& 60	& 60	& 60	& 50	& 30	& 20	& 10	\\
$\epsilon_{\mu\mu}$ [\%]
& 90	& 90	& 90	& 90	& 60	& 40	& 20
\\[1mm]\hline
\end{tabular*}
\caption{Run scenarios considered and effective efficiencies employed at the various centre-of-mass energies for the reconstructions of pairs of top quarks, bottom quarks and muons. When two beam polarizations are available, luminosities at each centre-of-mass energy are equally shared between them.}
\label{tab:run_scenarios}
\end{table}

For top-quark pair production, we heavily rely on the analysis of Refs.\,\cite{Durieux:2018tev, Durieux:2017gxd}. A linear effective-field-theory expansion is used throughout and was shown to be accurate.
So-called statistically optimal observables~\cite{Atwood:1991ka, Diehl:1993br} are defined on the resonant $\eett\to\bwbw$ final state. By construction, they maximally exploit the information contained in the total rate and differential distribution to extract the tightest constraints in the multidimensional space of operator coefficients.
Given a phase-space distribution
\newcommand{\dps}[1][]{\frac{\text{d}\sigma#1}{\text{d}\Phi}}
\begin{equation*}
\dps = \dps[_0] + C_i \dps[_i]
\end{equation*}
which depends linearly on small parameters $C_i$, observables statistically optimal for the determination of these $C_i$ at the $\{C_i=0,\,\forall i\}$ point are  the average values of $n\dps[_i]\big/\dps[_0]$ over the $n$ events collected~\cite{Atwood:1991ka, Diehl:1993br}.
In each collider run, the covariance matrix obtained through their ideal measurements is conveniently given by the phase-space integral
\begin{equation*}
\cov(C_i,C_j)^{-1} = \epsilon\mathcal{L}\int\!\text{d}\Phi
	\left(\dps[_i]\dps[_j]\bigg/\dps[_0]\right)
\end{equation*}
where $\mathcal{L}$ is the integrated luminosity of the run and $\epsilon$ is an  efficiency that can effectively account for acceptance, selection, resolution and reconstruction limitations.

The efficiencies we use for top-quark, bottom-quark and muon pair production at each centre-of-mass energy are also quoted in \autoref{tab:run_scenarios}. For top-quark pair production, full detector simulation studies have been carried out at CLIC centre-of-mass energies in semileptonic final states involving either a muon or an electron~\cite{Abramowicz:2018rjq}. Such a final state allows for an effective identification of the top-quark charges.
Fully hadronic final states could nevertheless be exploited in the future, and the more challenging reconstruction of final states involving a tau lepton could be tackled.
Being conservative, we however employ the effective efficiencies obtained in those full simulations, after average over beam polarization configurations.
Notable factors explaining the decrease of efficiency at higher centre-of-mass energies are the following. First, for top-quark pair production, the single top-quark production background becomes more significant and forces the use of more stringent selection cuts.
Second, the beam energy spectrum of linear colliders has a growing lower tail at higher energies. The effective luminosity actually collected close to the nominal energy is thereby reduced.
This motivates our choice of decreasing efficiencies for bottom-quark and muon pair production too.
We will comment below on the impact of the exact efficiencies assumed.

The constraints deriving from the measurements of $\eett\to\bwbw$ statistically optimal observables are presented in the Appendix~E of Ref.\,\cite{Durieux:2018tev}\footnote{See also the codes and numerics provided at \url{https://github.com/gdurieux/optimal_observables_ee2tt2bwbw}.} using the LHC TOP WG conventions~\cite{AguilarSaavedra:2018nen} which are directly related to the original Warsaw-basis definitions of Ref.\,\cite{Grzadkowski:2010es}.
Constraints on the CP-violating sector formed by the imaginary components of electroweak dipole operators are not considered.
We use the power counting estimates provided in \autoref{tab:coeffs} to convert those constraints to the $g_\star,m_\star$ parameter space of CH models. The cases of full $t$ and equal $q,t$ compositeness are both considered. We start by examining separately the impact of the various families of operators. We distinguish the following three categories: vertex operators $\qq{}{\phi q,\phi u}{}$, dipole operators $\qq{}{uW,uB}{}$ and four-fermion operators $\qq{}{lq,lu,eq,eu}{}$. This somewhat artificial distinction ---~it is basis-dependent~--- nevertheless permits to better understand from where the dominant constraint comes, in various regions of the $g_\star,m_\star$ plane.
We show in \autoref{fig:decomposition}, the sensitivities obtained for the CLIC-like, ILC-like, and CC-like benchmark run scenarios. Regions below the curves are probed at the one-sigma level. For simplicity, at this stage, the power-counting estimates for operator coefficients are assumed to be exactly satisfied.

Focusing first on top-quark pair production, it is seen that the constraints arising from four-fermion operators depend the most on the nature of the collider and, in particular, on the highest centre-of-mass energy available. At CLIC, they provide the dominant constraints (in both compositeness scenarios), except at values of $g_\star$ approaching $4\pi$ where vertex and dipole operators become marginally relevant.
Four-fermion constraints remain strong and flatten out as $g_\star$ increases in the case of a fully composite $t$. In that scenario, the four-fermion operator involving right-handed top quarks has a power counting estimate of the order of $g'^2/m_\star^2$ which is not suppressed by any negative power of $g_\star$.
At the ILC, and especially at circular colliders, vertex and dipole operators set the dominant constraints over somewhat larger ranges of $g_\star$ values. The effects of these operators do not significantly grow with the centre-of-mass energy. They are therefore better constrained with runs at lower energies, close to the peak of the top-quark production cross section, around $\sqrt{s}\simeq 400\,\gev$.
Black lines in \autoref{fig:decomposition} combine the constraints on all top-quark operators. Their non-vanishing correlations explain why this combination is sometimes less constraining than some categories of operators taken in isolation.

At circular colliders, note that the top-quark pair production threshold scan is usually employed to determine the top-quark mass (fixed here to $172.5\,\gev$), its width, and possibly the strong coupling constant.
Runs at two different energies are however required to constrain simultaneously the two- and four-fermion operators considered in Ref.\,\cite{Durieux:2018tev}.
It therefore remains to be examined whether $m_t$, $\Gamma_t$ and $\alpha_S(m_t)$ can be determined precisely together with all effective operator coefficients entering $\eett$, in such a CC-like run scenario.
Constraints can however be set in the two-dimensional $g_\star$,$m_\star$ parameter space with a run at $365\,\gev$ only. The resulting one-sigma limits on $m_\star$ are actually only loosened by a few percent.

\begin{figure}
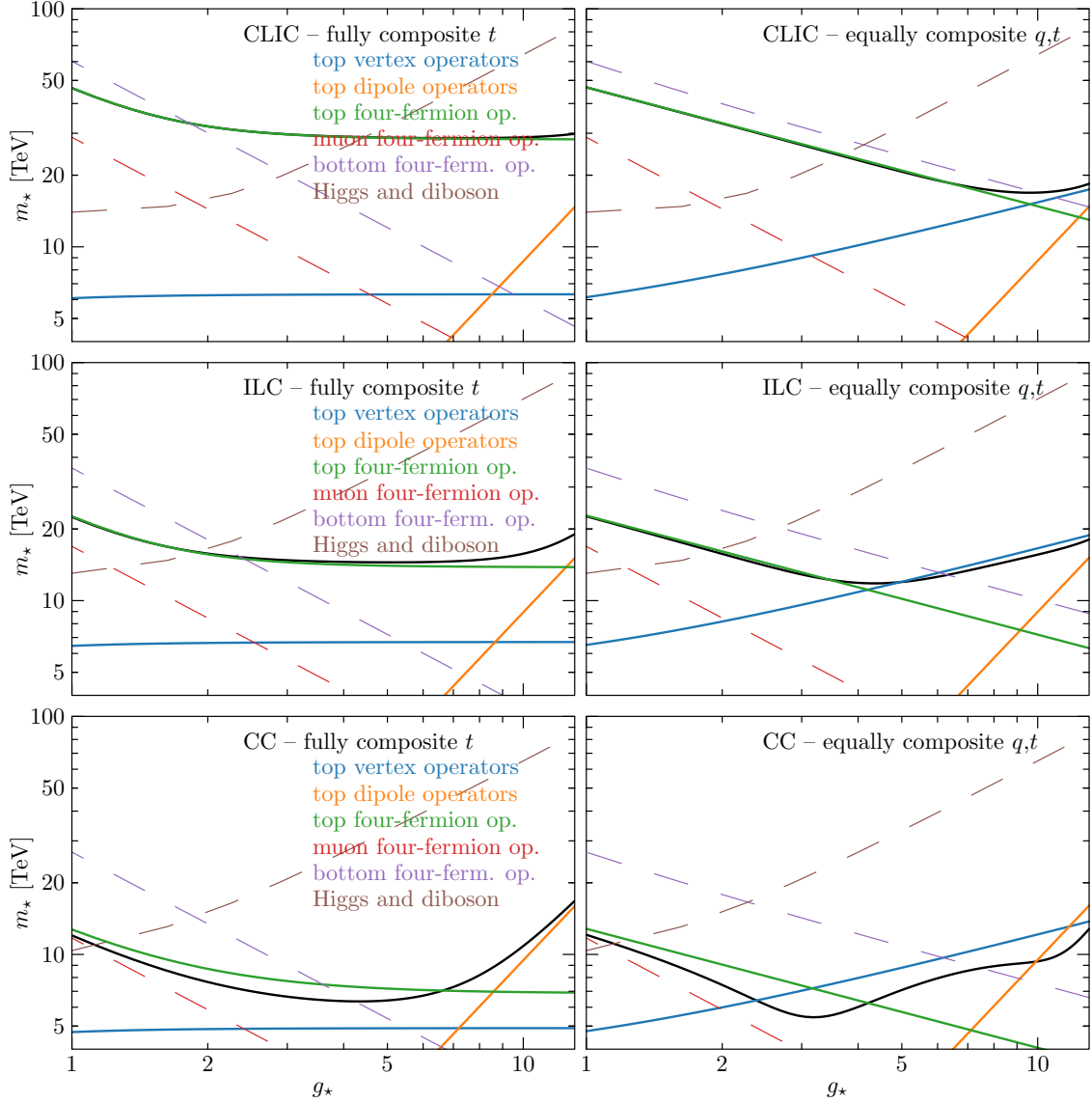

\includegraphics[trim= 0 19 0 0, clip, scale=.9]{CLIC_decomposition_full_1sigma_log.mps}%
\includegraphics[trim=25 19 0 0, clip, scale=.9]{CLIC_decomposition_partial_1sigma_log.mps}\\
\includegraphics[trim= 0 19 0 0, clip, scale=.9]{ILC_decomposition_full_1sigma_log.mps}%
\includegraphics[trim=25 19 0 0, clip, scale=.9]{ILC_decomposition_partial_1sigma_log.mps}\\
\includegraphics[trim= 0  0 0 0, clip, scale=.9]{CC_decomposition_full_1sigma_log.mps}%
\includegraphics[trim=25  0 0 0, clip, scale=.9]{CC_decomposition_partial_1sigma_log.mps}%
\caption{One-sigma sensitivities in the $g_\star,m_\star$ parameter space of CH models, deriving from statistically optimal observable measurements in top-quark pair production, when one single type of operators is considered at a time. The region below the black line is probed by top-quark production measurements, once all types of operators are combined.
The constraint arising from the universal contribution to four-fermion operators involving two electrons and two muons are displayed with a dashed red line.
The dashed violet line derives from constraints on four-fermion operators involving two electrons and two bottom quarks.
Higgs and diboson measurements~\cite{Durieux:2017rsg, DiVita:2017vrr} discussed below cover the region below the dashed brown curve.
For simplicity, the power counting of \autoref{tab:coeffs} is assumed to predict operator coefficients exactly.
}
\label{fig:decomposition}
\end{figure}

In addition to constraints arising from the measurements of statistically optimal observables in top-quark pair production, \autoref{fig:decomposition} also shows the ones which derive from the measurements of the statistically optimal observables relative to four-fermion operators in $\epem\to\mu^+\mu^-$ production (dashed red lines).
The $e^+e^-\mu^+\mu^-$ operators receive universal contributions from the strong sector, as discussed at the end of \autoref{sec:relevant_operators}.
In general, such limits are weaker than the ones arising from top-quark operators. Universal operators indeed have power-counting estimate of the order of $g^{(\prime)4}/g_\star^2m_\star^2$, and are thus suppressed with respect to top-quark operators which are of the order of $g^{(\prime)2}\epsilon_{t,q}^2/m_\star^2$.
Note that one could also include constraints arising from the production of any pair of charged lepton and quark. Roughly speaking this could improve our $\epem\to\mu^+\mu^-$ limit on $m_\star$ by a factor of about $\sqrt[4]{15}\simeq 1.9$.\footnote{Three flavours of charged leptons and four flavours of light quarks appearing in three colours would increase the statistics by a factor of about $15$. In the linear effective-field-theory regime, the limits on operators coefficients would improve by about $\sqrt{15}$. The operator coefficients being proportional to $1/m_\star^2$, the limits on $m_\star$ would improve by about $\sqrt[4]{15}$.}
As can be seen from \autoref{fig:decomposition}, such universal constraints could have an impact at low $g_\star$. For a more careful estimate, realistic reconstruction efficiencies should be evaluated.
At circular colliders, valuable universal constraints may also arise from the high-luminosity $Z$-pole run. A proper account of such measurements lies however beyond the scope of this paper.

In addition to universal contributions, four-fermion operators involving left-handed bottom quarks receive contributions enhanced by the mixing of the third-generation left-handed quark doublet $q$ to composite resonances.
We also use statistically optimal observables to simultaneously constrain the $e^+e^-b\,\bar{b}$ operators of vector Lorentz structure which interfere with standard-model amplitudes.
There are four of them when one accounts for the two possible chiralities of the two fermionic currents.
The optimal observable definitions are symmetrized between the $b$ and the $\bar{b}$ such that charge identification is not required (see Ref.\,\cite{Bilokin:2017lco} for a discussion).
The resulting one-sigma sensitivities in the $g_\star,m_\star$ plane are 	indicated with dashed violet lines in \autoref{fig:decomposition}.
In the case of a fully composite $t$, which minimizes the compositeness of $q$, the resulting constraints only surpass top-quark four-fermion ones for relatively small $g_\star$ (below about $2$ for CLIC- and ILC-like scenarios).
When the degree of compositeness of the left-handed third-generation doublet $q$ increases, as in our second scenario, $e^+e^-b\,\bar{b}$ operators provide more stringent constraints than $e^+e^-t\,\bar{t}$ ones over the whole range of acceptable $g_\star$.
The higher efficiencies in bottom quark reconstruction play a major role. Lower limits on $m_\star$ scale as $\sqrt[4]{\epsilon}$.
With identical efficiencies, constraints arising from top and bottom four-fermion operators would overlap almost perfectly in the equally composite $q,t$ scenario.
Note that momentum-independent modifications of the left-handed bottom-quark coupling to the $Z$ boson are suppressed due to the $P_{LR}$ symmetry mentioned in \autoref{sec:power_counting_estimates}.
One does therefore not expect the corresponding vertex operators to be more constraining than top-quark ones which dominate at high $g_\star$.

Constraints arising from Higgs and diboson measurements at future lepton colliders are displayed with dashed brown lines in \autoref{fig:decomposition}.
They are derived from the global effective-field-theory analysis performed in Refs.\,\cite{Durieux:2017rsg, DiVita:2017vrr}.
More details are provided in the next section and in \autoref{sec:higgs}.

\section{Discovery reach}
\label{sec:disc}

We finally derive the combined reach of top- and bottom-quark pair production on CH scenarios and compare it with that of Higgs and diboson measurements at future lepton colliders. We also discuss the interplay between such measurements and naturalness considerations.
For this purpose, it is useful to first introduce a new parameter:
\be
\Delta = f^2/v^2
\ee
given by the ratio of $f\equiv m_\star/g_\star$ (often called \emph{Goldstone decay constant}) to the Higgs vacuum expectation value $v$ squared.
It has two important interpretations. First, $\Delta$ measures the fine tuning of the Higgs potential (see Refs.\,\cite{Matsedonskyi:2012ym,Panico:2012uw} for detailed discussions). A generic non-tuned estimate for the Higgs vacuum expectation value in CH models is $v \sim f$.
However, $\Delta$ has to be significantly larger than one to satisfy the experimental constraints which push $f$ to \tev\ scales.
The currently preferred value, $\Delta \sim 10$, does not seem too fine-tuned. Minimal CH models would however become unappealing with $\Delta$ one or two orders of magnitude larger.
Less minimal constructions, like that of Refs.\,\cite{Csaki:2017eio, Batell:2017kho}, can however accommodate large values of $\Delta$ at the price of an increased model complexity. It is therefore not unreasonable to consider the region of parameter space with $\Delta\gg 1$.
Second, $1/\Delta$ directly controls the deviations in Higgs couplings with respect to SM predictions. We will get back to these effects at the end of this section.
Deviations from the SM generically vanish in the limit of large tuning.\footnote{Such a large tuning can however have an impact on the cosmological evolution of the universe and therefore have detectable signatures~\cite{Amin:2018kkg, Bruggisser:2018mrt, Bruggisser:2018mus}.}

\begin{figure}[!t]
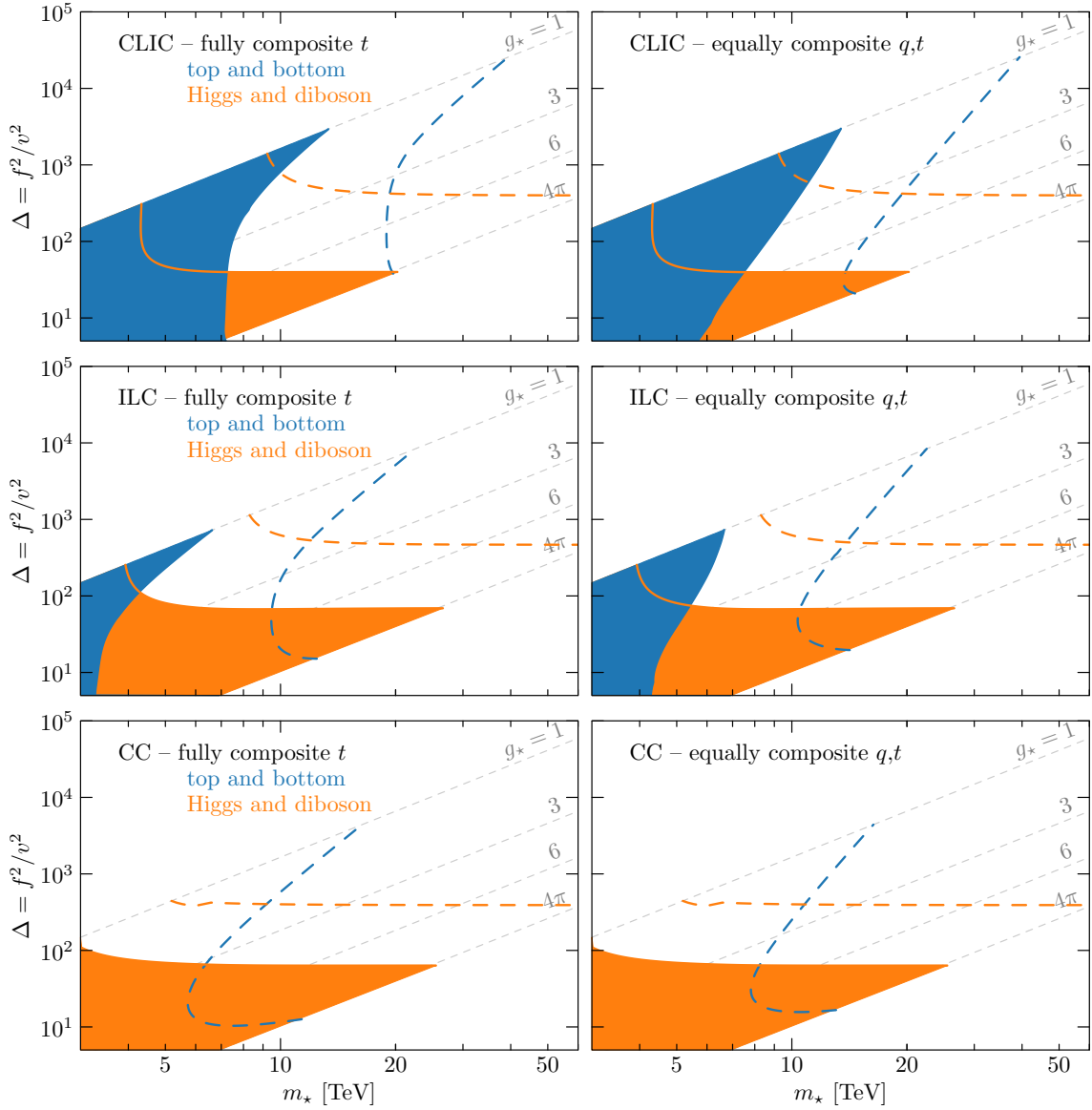
\centering
\includegraphics[trim=0 22 0 0, clip, scale=.9]{CLIC_xi_full_5sigma_floatNflip_comb.mps}%
\includegraphics[trim=28 22 0 0, clip, scale=.9]{CLIC_xi_partial_5sigma_floatNflip_comb.mps}\\
\includegraphics[trim=0 22 0 0, clip, scale=.9]{ILC_xi_full_5sigma_floatNflip_comb.mps}%
\includegraphics[trim=28 22 0 0, clip, scale=.9]{ILC_xi_partial_5sigma_floatNflip_comb.mps}\\
\includegraphics[trim=0 0 0 0, clip, scale=.9]{CC_xi_full_5sigma_floatNflip_comb.mps}%
\includegraphics[trim=28 0 0 0, clip, scale=.9]{CC_xi_partial_5sigma_floatNflip_comb.mps}%
\caption{Five-sigma discovery reach, at future linear and circular lepton colliders, in the CH model parameter space characterized by $m_\star$ and $\Delta$. The coupling $g_\star$ takes constant values on the dashed grey contours. Values below $1$ and above $4\pi$ are not allowed.
The regions probed by top- and bottom-quark pair production measurements are indicated in blue. The orange regions are accessible through the Higgs and diboson measurements.
Solid and dashed lines respectively delimit regions probed in pessimistic and optimistic cases.
These extremes are obtained by assuming the power-counting estimates of operator coefficients are satisfied up to factors ranging between $1/2$ and $2$, considering all possible combinations of relative signs between them.
}
\label{fig:limits_all}
\end{figure}

The five-sigma discovery reaches brought by the combination of third-generation-quark pair production measurements at lepton colliders are displayed as blue contours in \autoref{fig:limits_all}.
The pessimistic (solid) and optimistic (dashed) reaches are obtained by assuming that the power-counting estimates of operator coefficients are satisfied up to independent factors varying from $1/2$ to $2$.
All possible combinations of relative signs are also considered to cover mutual cancellations and enhancements.
Such measurements, in a CLIC-like run scenario, would ensure the discovery of models with $m_\star\lesssim 5\,\tev$ and $\Delta\lesssim 3$. Their maximal reach would extend to $m_\star \sim 40\,\tev$ mass scales and intolerable tunings of the order of $\Delta \sim 10^4$. Overall, the sensitivity to $m_\star$ and $\Delta$ increases for lower $g_\star$. The sizeable difference between the two compositeness scenarios at higher $g_\star$ values due to different scalings for four-fermion operators featuring a pair of right-handed top quarks was already noted in the previous section.
As seen there too, the lower centre-of-mass energies accessible in an ILC-like scenario reflects in a weaker sensitivity to the CH parameter space.
The top-portal effects driven by mixings of the top quark to composite partners which dominate at high energies can be modelled with four-fermion operators. This weaker sensitivity at lower centre-of-mass energies is even clearer in a CC-like scenario.
Large cancellations between different contributions are possible and the pessimistic discovery reach no longer remains relevant.

For the sake of comparison, let us also briefly discuss the reach of the Higgs and diboson measurements at future lepton colliders in the CH parameter space.
They dominantly probe the universal effects of compositeness that do not depend on the SM fermion mixings with composite resonances.
We use the prospective sensitivities derived in Refs.\,\cite{Durieux:2017rsg, DiVita:2017vrr}. More details are given in \autoref{sec:higgs}.
The list of relevant SILH basis operators is given in \autoref{tab:higgs}, together with the corresponding power-counting estimates of their coefficients.
For $g_\star$ larger than about $2$, the overall reach is dominated by the constraints on $O_H$ and $O_b$ operator coefficients.
The latter modifies the dominant $h\to\bbbar$ branching fraction while the former gives universal contribution to all Higgs production and decay processes.
The power counting estimates for their coefficients are $1/v^2 \Delta$ and  $\lambda_b/v^2 \Delta$.
Their magnitudes are thus controlled solely by the degree of tuning.
The orange contours delimiting the regions probed by Higgs measurements in the $m_\star,\Delta$ plane of \autoref{fig:limits_all} therefore become horizontal as $g_\star$ increases.
Again, regions below the solid and dashed contours are respectively probed in pessimistic and optimistic cases.
The constraints on $c_W-c_B$ become relevant for smaller $g_\star$. Constraints on the top-quark and tau-lepton Yukawa operators are only marginally relevant.
The ones which apply on the Higgs trilinear self-coupling are too weak to have any impact.
Unlike top-portal effects whose discovery reach increases with the available centre-of-mass energy, universal ones bring similar sensitivities in all three benchmark scenarios considered.

The top-portal and universal probes are very much complementary.
Universal effects sensed via Higgs measurements efficiently cover the parameter-space regions of lowest tuning.
The top portal effects which manifest themselves in \eett\ and \bbbar\ productions help covering lower $g_\star$ coupling values and push the discovery reach on the compositeness mass scale deeper into the multi-\tev\ region.
A combination of both types of measurements therefore covers efficiently the range of acceptable couplings, the most natural parameter space, and large mass scales.

\section{Conclusions}

Composite Higgs (CH) models motivated by the gauge hierarchy problem predict new physics in the vicinity of the \tev\ scale.
In these models, the sizeable top-quark mixings with composite partners are required to generate a top-quark Yukawa of order one.
We examined the prospects of discovering Higgs compositeness at future lepton colliders through this top portal in precision measurements of top- and bottom-quark pair production.

The energy-growing top-portal effects can be modelled by four-fermion operators.
With third-generation-quark pair production at \tev\ centre-of-mass energies, linear colliders probe mass scales much higher than the direct discovery reach of the LHC.
At the lower energies accessible with circular machines, Higgs compositeness is more likely to manifest itself through universal effects.
Measurements of Higgs couplings then have a more robust constraining power.
Both types of measurements exhibit complementary discovery reaches in the CH parameter space.

Our main observation is that a significant fraction of this parameter space can be covered by future linear lepton colliders with top- and bottom-quark pair production measurements only.
With centre-of-mass energies in the multi-\tev\ range, CLIC would for instance conservatively discover new composite dynamics with mass below about $5\,\tev$.
It would moreover have chances of discovering compositeness mass scales as high as $40\,\tev$.
 
\subsubsection*{Acknowledgements}

We are grateful to Andrea Wulzer, Giuliano Panico and Christophe Grojean for useful discussions, and to Jiayin Gu for help with the Higgs measurement likelihoods of Refs.\,\cite{Durieux:2017rsg, DiVita:2017vrr}.

\appendix
\section{Higgs and diboson measurements}
\label{sec:higgs}

The composite Higgs interpretation of the Higgs and diboson measurement prospects presented in Refs.\,\cite{Durieux:2017rsg, DiVita:2017vrr} is detailed here. The likelihood obtained in a subset of the SILH basis~\cite{Giudice:2007fh} is employed (see Appendix~A of Ref.\,\cite{Durieux:2017rsg}).
CP-conservation and perfect electroweak precision measurements are assumed there, $c_W+c_B=0$ in particular.
Departures from flavour universality are only allowed to distinguish the various modifications of fermion Yukawa couplings: for the top, bottom and charm quarks, muon and tau leptons.
Double Higgs production as well as the loop-level dependence of single Higgs production and decay modes on the Higgs trilinear self-coupling are included.
The power counting we adopt only differs from that of Ref.\,\cite{Giudice:2007fh} for $c_\gamma$ by the loop suppression factor which we take as $\lambda_t N_c/16\pi^2$ with $N_c=3$ the number of colours, instead of $g^2/16\pi^2$.
The operators, their normalization and the power counting estimates used are provided in \autoref{tab:higgs}.

The CLIC, ILC and FCC-ee run scenarios of Ref.\,\cite{DiVita:2017vrr} are employed, with integrated luminosity equally split between two beam polarizations, when available:
\begin{center}
\begin{tabular*}{.6\textwidth}{@{\extracolsep{\fill}}lccc}
\hline\noalign{\vskip.5ex}
	& $\sqrt{s}$ [\gev]	& $\mathcal{L}$ [\iab]	& $P(e^+,e^-)$
\\[.5ex]\hline\noalign{\vskip.5ex}
FCC-ee	& $240$			& $5$			& $(0,0)$		\\
	& $350$			& $1.5$			& $(0,0)$		\\\hline\noalign{\vskip.5ex}
ILC	& $250$			& $2$			& $(\pm0.3,\mp0.8)$	\\
	& $350$			& $0.2$			& $(\pm0.3,\mp0.8)$	\\
	& $500$			& $4$			& $(\pm0.3,\mp0.8)$	\\
	& $1000$		& $2$			& $(+0.2,-0.8)$		\\\hline\noalign{\vskip.5ex}
CLIC	& $350$			& $0.5$			& $(0,0)$		\\
	& $1400$		& $1.5$			& $(0,0)$		\\
	& $3000$		& $2$			& $(0,0)$
\\[.5ex]\hline
\end{tabular*}
\end{center}
They differ mildly from the ones we adopted for top-quark pair production measurements. The constraints in the CH parameter space however have a very mild dependence on the run scenario considered. They are indeed largely dominated by the limit applying on $c_H$ which is remarkably universal. A  prospective $c_H\lesssim 0.002$ individual constraint applies at the one-sigma level for all three collider run scenarios.
With a power counting estimate for $c_H$ given by $g_\star^2v^2/m_\star^2$ (see \autoref{tab:higgs}), one then approximately obtains $m_\star\gtrsim g_\star v/\sqrt{0.002}$ (e.g., $m_\star\gtrsim 30\,\tev$ for $g_\star=5$, still at the one-sigma level).

See Sec.\,6.2 of Ref.\,\cite{Gu:2017ckc} for a similar CH interpretation of Higgs measurements at future lepton colliders.

\begin{table}[h]
\hfil\vbox{\halign{#&\hfill$\displaystyle#$\hfill\quad&\quad\hfill$\displaystyle#$\hfill\quad&\quad\hfill$\displaystyle#$\hfill\cr
\noalign{\hrule\vskip1ex}
&\omit \hfill coefficient	\hfill
&\omit \hfill operator		\hfill
&\omit \hfill power counting	\hfill
\cr
\noalign{\vskip1ex\hrule\vskip1ex}
&c_W \frac{g}{m_W^2}
	& \frac{1}{2} (\phi^\dagger i\overleftrightarrow{D}_{\mu}^a \phi)
		\: D^\nu W^a_{\mu\nu}
	& \frac{g}{m_\star^2}
\cr
&c_B \frac{g'}{m_W^2}
	& \frac{1}{2} (\phi^\dagger i\overleftrightarrow{D}_{\mu} \phi)
		\: \partial^\nu B_{\mu\nu}
	& \frac{g'}{m_\star^2}
\cr
&c_\gamma \frac{g'^2}{m_W^2}
	& \phi^\dagger\phi
		\: B_{\mu\nu}B^{\mu\nu}
	& \frac{g'^2}{m_\star^2} \frac{\lambda_t^2}{16\pi^2} N_c
\cr
&c_{HB} \frac{g'}{m_W^2}
	& i(D^\mu\phi)^\dagger (D^\nu\phi)B_{\mu\nu}
	& \frac{g'}{m_\star^2}\frac{g_\star^2}{16\pi^2}
\cr
&c_{HW} \frac{g}{m_W^2}
	& i (D^\mu\phi)^\dagger \tau^a (D^\nu\phi)W^a_{\mu\nu}
	& \frac{g}{m_\star^2} \frac{g_\star^2}{16\pi^2}
\cr
&c_{H} \frac{1}{v^2}
	& \frac{1}{2} (\partial_\mu |\phi|^2)^2
	& \frac{g_\star^2}{m_\star^2}
\cr
&c_{g} \frac{g_S^2}{m_W^2}
	& |\phi|^2 G^A_{\mu\nu} G^{A\mu\nu}
	& \frac{g_S^2}{m_\star^2} \frac{\lambda_t^2}{16\pi^2}
\cr
&c_u \frac{\lambda_u}{v^2}
	& \bar{q} \tilde{\phi} u\: |\phi|^2
	& \frac{\lambda_u g_\star^2}{m_\star^2}
\cr
&c_d \frac{\lambda_d}{v^2}
	& \bar{q} \phi d\: |\phi|^2
	& \frac{\lambda_d g_\star^2}{m_\star^2}
\cr
&c_e \frac{\lambda_e}{v^2}
	& \bar{l} \phi e\: |\phi|^2
	& \frac{\lambda_e g_\star^2}{m_\star^2}
\cr
&c_6 \frac{\lambda}{v^2}
	& -|\phi|^6
	& \frac{g_\star^4}{m_\star^2} \frac{\lambda}{g_\star^2}
\cr
&c_{3W} \frac{g}{3!m_W^2}
	& \epsilon_{abc} W^{a\nu}_\mu W^{b}_{\nu\rho}W^{c\rho\mu}
	& \frac{g^3}{g_\star^2m_\star^2} \frac{g_\star^2}{16\pi^2}
\cr
\noalign{\vskip1ex\hrule}}}
\caption{SILH operators, their coefficients in Refs.\,\cite{Durieux:2017rsg, DiVita:2017vrr}, and the power counting adopted here for the latter.}
\label{tab:higgs}
\end{table}

\bibliographystyle{apsrev4-1_title}
\bibliography{ops}
\end{document}